\documentclass[prl,twocolumn]{revtex4}
\usepackage[cp1251]{inputenc}
\usepackage{amsmath}
\usepackage{amssymb}
\usepackage{euscript}
\usepackage[dvips]{graphicx}
\usepackage{epsfig}
\usepackage[usenames,dvipsnames]{color}
\usepackage{longtable}
\usepackage{ulem}
\usepackage{txfonts}
\usepackage{bm}

\usepackage{aas_macros}

{}

\newcommand{\pot}[2]{#1\times10^{#2}}{}
\newcommand{\expf}[1]{{\rm e}^{#1}}{}
\newcommand{\Planck}{{\it Planck}}{}

\usepackage{color}

\begin{document}
%\title[Dipoles of CMB distortions]
%{Dipole anisotropy of relic CMB distortions}

\title[Distortions of the CMB dipole]
{Spectral distortions of the CMB dipole}

\author{S.A.~Balashev$^{1,2}$}
\author{E.E.~Kholupenko$^{1}$}
\author{J.~Chluba$^{3, 4}$}
\author{A.V.~Ivanchik$^{1,2}$}
\author{D.A.~Varshalovich$^{1,2}$}
\affiliation{$^1$Ioffe Institute, {Polytekhnicheskaya 26}, 194021 Saint-Petersburg, Russia;}
\email{s.balashev@gmail.com}
\affiliation{$^2$Peter The Great St.Petersburg Polytechnic University, {Polytekhnicheskaya 29}, 195251 Saint-Petersburg, Russia}
\affiliation{$^3$Kavli Institute for Cosmology Cambridge, Madingley Road, Cambridge, CB3 0HA, UK}
\affiliation{$^4$Department of Physics and Astronomy, Johns Hopkins University, 
3400 N. Charles Steet, Baltimore, MD 21218, USA}

\date{\today}

\begin{abstract}
We consider the distortions of the cosmic microwave background (CMB) dipole anisotropy related to the primordial recombination radiation (PRR) and primordial $y$- and $\mu$-distortions. The signals arise due to our motion relative to the CMB restframe and appear as a frequency-dependent distortion of the CMB temperature dipole. To leading order, the expected relative distortion of the CMB dipole does not depend on the particular observation directions and reaches {the level of} $10^{-6}$ for the PRR- and $\mu$-distortions and $10^{-5}$ for the $y$-distortion in the frequency range 1 -- 700~GHz. The temperature differences arising from the dipole anisotropy of the relic CMB distortions depend on the observation directions. For mutually opposite directions, collinear to the CMB dipole axis, the temperature differences because of the PRR- and $\mu$-dipole anisotropy attain values $\Delta T\simeq 10\,$nK in the considered range. The temperature difference arising from the $y$-dipole anisotropy {may reach} values of up to {$1\,\mu$K}. The key features of the considered effect are as follow: (i) an observation of the effect does not require absolute calibration; (ii) patches of sky with minimal foreground contamination can be chosen. Future measurements of the CMB dipole distortion thus will provide an alternative method for direct detection of the PRR-, $y$-, and $\mu$-distortions. The $y$-distortion dipole may be detectable with PIXIE at a few standard deviations.
\end{abstract}


\maketitle

\section{Introduction}
\label{introduction}
\noindent
Observations of the cosmic microwave background (CMB) temperature and polarization anisotropies constitute one of the major pillars of the standard cosmological model. The CMB power spectra provide an opportunity to determine the cosmological parameters of the Universe, since these are mainly determined by cosmological recombination \citep{Zeldovich1968, Peebles1968}, acoustic oscillations \citep{Sunyaev1970, Peebles1970}, diffusion damping \citep{Silk1968} and other well-studied processes in the early Universe \citep{Hu2002}. Measurements taken with the {\sl Wilkinson Microwave Anisotropy Probe} (WMAP) allow us to determine most of the cosmological parameters with a precision better than 4\% \citep{Hinshaw2013}. \Planck\ improved these measurements, reaching an accuracy level of $\simeq 1-2$\%  \cite{Plank2015Cosmo}. While cosmological constraints are driven by anisotropies at multipoles $\ell>1$, the largest CMB anisotropy is caused by our motion relative to the CMB restframe, manifesting itself as a temperature dipole \cite[e.g.,][]{Conklin1969, Fixsen1994}. 

It has been shown that the average CMB spectrum is Planckian with a relative accuracy better than $\simeq 10^{-3}$ \cite{Fixsen1996, Mather1999, Fixsen2009}. However, it is well known that tiny deviations of the CMB from the Planckian spectrum are expected {\citep[e.g.,][]{Zeldovich1968, Peebles1968, Sunyaev_Zeldovich1969, Sunyaev1970mu, Illarionov1975, Burigana1991, Hu1993, Chluba2011therm, Sunyaev2013, Tashiro2014}}. 
One inevitable distortion was formed as a result of the uncompensated transitions of recombining electrons during the cosmological recombination era at redshifts $z\simeq$ 800 -- 8000 \citep{Dubrovich1975, Bernshtein1977, RybickiDell94, Dubrovich_Stolyarov1995, Dubrovich_Stolyarov1997, Spaans_Norman1997, Kholupenko2005, RubinoMartin2006, Sunyaev2009}. The other expected distortions are the primordial $\mu$- and $y$-distortions \citep{Sunyaev_Zeldovich1969, Sunyaev1970mu} caused by episodes of early energy release. Here, we argue that the spectrum of the CMB dipole could be used to constrain and detect these signals. 

The physics of the primordial recombination process are very clear and has been studied in detail \cite[e.g.,][]{Grachev_Dubrovich1991, Seager2000, Dubrovich2005, Chluba2006twophoton, Kholu2006, Kholupenko2007, Grachev_Dubrovich2008, Switzer2008, Switzer_Hirata2008b, Hirata_Switzer2008, Hirata2008, Grin_Hirata2010, AliHaimoud_Hirata2011}. 
We will refer to the total spectrum from recombining electrons and ions as the primordial recombination radiation (PRR). 
The number of CMB photons is $\simeq 10^9$ times larger than the number of baryons. Thus, the distortion of the CMB spectrum is expected to be $\simeq 10^{-9}$ -- $10^{-8}$ of the total CMB energy density (it was shown that about 5 photons per hydrogen atom were released as the  PRR spectrum; \citealt{Chluba2006}). However, it was found that in some frequency range these distortions can exceed $10^{-7}$ of the CMB spectrum \cite{RubinoMartin2006, Chluba2006}. The PRR spectrum mainly consists of the Lyman, Balmer, Paschen, Brackett and other series \citep{Chluba2006, Chluba2007}. Also, there are contributions from various processes, such as 2s-1s two-photon decay (\citealt{Chluba2008, Hirata2008}; which gives about 50\% of the emission around 1\,THz) and free-bound emission (\citealt{Chluba2006}; about 20-30\% at the whole frequency range). In addition, it was found that He\,{\sc ii} and He\,{\sc i} recombination spectra contribute about 10\%-15\%, on average, to the PRR and up to 50\% at specific frequencies \citep{RubinoMartin2008, Chluba2009c}. 

Measuring of the PRR spectrum is very important for the several reasons: (i) it will allow us to measure the CMB monopole temperature $T_0$ \citep{Chluba2008b}; (ii) it can be used to estimate cosmological parameters like the fraction of baryon matter, $\Omega_{\rm b}$, and the helium to hydrogen abundance ratio, $Y_{\rm p}$ \citep{Chluba2008b}; 
(iii) additional energy release (e.g., annihilation or decay of dark matter particles) can be constrained \citep[e.g.,][]{Chluba2009,Dubrovich_Grachev2009,Chluba2010,Grachev_Dubrovich2011}, since the PRR formed before the epoch of last scattering (especially the PRR components arising from He\,{\sc ii} and He\,{\sc i} recombinations); and (iv) variations of fundamental constants could potentially be probed \citep{Chluba2010}. In addition, the non-detection of the PRR spectrum at the expected level and with the predicted shape would be a serious challenge for the standard cosmological model.

To create a $\mu$-distortion \citep{Sunyaev1970mu} requires efficient energy exchange between matter and photons, so that this type of distortion is {\it only} formed at redshifts $z\gtrsim \pot{5}{4}$, while at lower redshifts a $y$-type distortion, also known in connection with the thermal Sunyaev-Zeldovich (SZ) effect \citep{Sunyaev_Zeldovich1969}, is created \citep{Burigana1991, Hu1993}. The amplitude of these signals is more uncertain, but even within our standard cosmological paradigm, one expects an average $y$-parameter of $y\simeq 10^{-7}-10^{-6}$ due to the large-scale structure and the reionization epoch \citep[e.g.,][]{Hu1994pert, Cen1999, Refregier2000, Oh2003}, with the most recent computations giving $y\simeq\pot{2}{-6}$ \citep{Hill2015}. The primary contribution to the $y$-distortion is from galaxy groups and clusters after reionization, while the contribution from the reionization era and the intergalactic medium is $y\simeq \pot{2}{-7}$ \citep{Hill2015}. For the $\mu$-distortion, the dissipation of small-scale acoustic modes {\citep[e.g.,][]{Sunyaev1970mu, Daly1991, Hu1994}} in the standard slow-roll inflation scenario is expected to give rise to $\mu\simeq \pot{2}{-8}$ \citep{Chluba2012}. By measuring the $\mu$-distortion, we can learn about the small-scale power spectrum at wavenumbers $1\,{\rm Mpc}^{-1}\lesssim k\lesssim 10^4\,{\rm Mpc}^{-1}$, and thus constrain different early-universe models \citep{Chluba2012inflaton, Khatri2013forecast, Chluba2013PCA, Clesse2014}.

To detect the average $\mu$- and $y$-distortion, an {\it absolute calibration} and measurements in wide bands at high frequencies ($\nu\simeq 30\,$GHz -- 1\,THz) are required. The spectral shapes of these signals are very broad and therefore measurements are more challenging. Experimental concepts like PIXIE \citep{Kogut2011PIXIE} and PRISM \citep{PRISM2013WPII} may reach the required sensitivity and stability to detect the aforementioned signals. In contrast, for the PRR, one can also make use of its unique spectral dependence \citep{Sunyaev2009}, so that absolute calibration in frequency is not necessarily required, but one merely needs a sufficient {\it inter-channel calibration} to extract the typical peak-to-peak amplitude of $\simeq10$\,nK at low frequencies. In the future, this type of measurement may even be possible at low frequencies  from the ground \citep{SathyanarayanaRao_et_al2015}, but an improved version of PIXIE could also succeed \citep{Desjacques2015}. For these observations, the flux can be collected from large regions of the sky which are minimally contaminated by the Galaxy and other backgrounds. In addition, one can make use of the fact that the distortion signals should be unpolarized \citep{Sunyaev2009}.

\citet{Dubrovich2004} noted that the PRR signal becomes much more prominent relative to the CMB if one observes the differential PRR spectrum and compares it with the differential spectrum of CMB. To measure derivatives, it is necessary to subtract the spectrum at adjacent frequencies. In principle, this also requires absolute calibration for different frequency channels; however, the derivative of the average spectrum can alternatively be obtained by measuring the CMB dipole spectrum. 
Recently, it was shown \citep{Kholupenko2015} that the SZ effect, which introduces a shift and distortions of the CMB spectrum through Comptonization, can also be used to calculate the PRR spectrum derivative. However, it is difficult to observe this effect with modern equipment because the expected signal is extremely weak due to the small collection area (it is determined by the angular size of the galaxy cluster chosen for observation). Additionally, this effect is model-dependent. It demands the determination of the galaxy cluster model for which the SZ effect observation is performed. 

In this paper, we discuss the detailed spectrum of the CMB dipole anisotropy arising due to the motion of an observer relative to the CMB restframe. Previously, this effect was considered by \cite{Danese1981} and \cite{Smoot1997}, but detailed calculations and numerical estimates considering modern technology and the modern cosmological model were not performed. The effect can be used as a natural method to measure the derivatives of the PRR spectrum and the $\mu$- and $y$-distortions, but also extragalactic (non-comoving) foreground signals. It should be emphasized that measurements of this effect do not require absolute calibration. We show that the motion of the Solar System relative to the CMB restframe is enough to consider the effect for future experiments.

\section{Spectrum of the CMB dipole}
\label{dipole}
\noindent
It is well known that relative to the CMB restframe, the Solar System is moving toward ($l,b$) = (263.99°$\pm$0.14°, 48.26°$\pm$0.03°) \citep{Fixsen1996,Hinshaw2009} in galactic coordinates. The speed of this motion is $\varv=369.0\pm0.9$ km/s and produces the CMB dipole anisotropy due to the Doppler effect. Using the CMB dipole, the first derivative of the monopole spectrum can be calculated. To do this, one can subtract the spectra obtained from the different directions using two identical instruments, a procedure that directly cancels the dominant CMB monopole.

Denoting the total occupation number of the monopole spectrum as $\eta_{\rm m}(\nu)$, at lowest order in $\beta=\varv/c$, the associated motion-induced dipole spectrum is given by 
\begin{equation}
\eta_{\rm d}(\nu, \Theta)\approx \eta_{\rm d}(\nu) \cos \Theta,
\label{dipole_spectrum}
\end{equation}
where {the} angle $\Theta$ is measured with respect to the motion direction, and 
the occupation number amplitude is given {by}
\begin{equation}
\eta_{\rm d}(\nu)\approx -\nu\partial_\nu \eta_{\rm m}(\nu)\,\beta.
\label{dipole_spectrum_amplitude}
\end{equation}
Assuming that the CMB monopole spectrum is just a blackbody with $\eta_{\rm CMB, m}=1/(\expf{x}-1)$, where $x=h\nu /\left( k_{B}T_{0}\right)$ is the dimensionless frequency, we obtain the well-known CMB temperature dipole
%---------------------
\begin{equation}
{\eta_{\rm CMB, d}(\nu)}=\beta G(x),~~G(x)= \frac{x\,\expf{x}}{(\expf{x}-1)^2}, 
\label{T_dipole_spectrum}
\end{equation}
%---------------------
where {the} function $G(x)$ exhibits the spectrum of thermal fluctuations. Note that $\nu$ is the measured frequency in the observer frame. The motion-induced temperature dipole causes the dominant contribution to the dipole anisotropy. 

When calculating the contributions from the PRR, $\mu$-, and $y$-distortions to the CMB dipole spectrum, we should specify the spectral shapes of these distortions. For the PRR spectrum, we use the results given by \citet{RubinoMartin2008}, with digital data taken from personal web page of J.A.~{Rubi{\~n}o-Mart{\'{\i}}n} at Istituto de Astrof{\'{\i}sica de Canarias site. \footnote{http://www.iac.es/galeria/jalberto/pages/research-interests/cosmological-recombination.php}  Since no simple analytic approximation for the PRR exists, we numerically evaluate the frequency derivative in Eq.~\eqref{dipole_spectrum_amplitude}. The associated occupation number amplitude is denoted as $\eta_{\rm PRR,d}(\nu)$.

For the $y$-distortion, the average occupation number was calculated by the following formulae \citep{Zeldovich_Sunyaev1969}: 
\begin{equation}
\eta_{y, \rm m}(\nu)=y Y(x),~~Y(x)=G(x)\left[x\coth(x/2)-4\right]{.}
\end{equation} 
{Therefore,} for the motion-induced dipole signal related to the average $y$-distortion, we have
%---------------------
\begin{equation}
\eta_{y, \rm d}(\nu)\approx - \beta y \left[ 20 G(x) + 6 Y(x) - x^2G(x)-6 x G(x)^2\right],
\label{y_distortion_def}
\end{equation}
%---------------------
which was determined by simply evaluating $\nu\partial_\nu \eta_{y, {\rm m}}\!=\!x\partial_x Y(x)$. We will use a fiducial value of {$y=\pot{2}{-6}$ \citep{Hill2015} as an estimate, neglecting the relativistic temperature corrections \citep{Itoh98, Sazonov1998, Challinor1998, ChlubaSZpack} caused by hotter group-size systems. 

For the $\mu$-distortion, the average occupation number is
%---------------------
\begin{equation}
\eta_{\mu, {\rm m}}=\frac{1}{\expf{x\phi +\mu}-1}-\frac{1}{\expf{x}-1}\approx \mu M(x),
\label{muav_distortion_def}
\end{equation}
%---------------------
where the function $\phi (\mu)\approx 1-0.4561\mu$ is the temperature correction providing the total photon number density conservation and {the} function $M(x)$ is 
\begin{equation}
M(x)=G(x)\!\left[0.4561-x^{-1}\right]{.}
\end{equation}
Thus, the related distortion to the dipole is given by
%---------------------
\begin{equation}
\eta_{\mu, \rm d}(\nu)
=\beta\mu\left[\left(3+\frac{Y(x)}{G(x)}\right)M(x)-\frac{G(x)}{x}\right].
\label{mu_distortion_def}
\end{equation}
%---------------------
We will use $\mu=\pot{2}{-8}$ as a fiducial value. 
In the following, we neglect any corrections from free-free distortion at low frequencies due to reionization \citep[e.g.,][]{Burigana2004, Trombetti2014} and foregrounds, which do not constitute primordial signals but could also be directly constrained using measurements of the dipole spectrum. The difference between galactic (comoving) and extragalactic (non-comoving) signals may provide an additional handle for component separation of the monopole signals. We also omit corrections to the shapes of the $\mu$- and $y$-distortion caused by the thermalization process at frequencies $\nu\simeq 1-10\,$GHz \citep{Burigana1991, Hu1993, Chluba2011therm}. Furthermore, we assume that the $y$-distortion of the dipole caused by the aberration and boosting \citep{Challinor2002, Planck2013abber} of the CMB temperature quadrupole, an artifact of the map-making procedure \citep{Chluba2004, Dai2014, Quartin2015}, is separated.

\section{Importance of the effects}
To illustrate the effects, let us define a measure of the relative anisotropy by
\begin{equation}
d\left(\nu, \bm{n_2},\bm{n_1}\right)={\eta \left(\nu, \bm{n_2}\right) \over \eta \left(\nu, \bm{n_1}\right) } - 1,
\label{d_Definition}
\end{equation}
where $\eta(\nu, \bm{n})$ is the total photon occupation number in the direction $\bm{n}$. The dominant contribution to $d\left(\nu, \bm{n_2}, \bm{n_1}\right)$ is due to the CMB temperature dipole itself and can be expressed as 
%------------------------
\begin{equation}
d_{\rm CMB}(\nu,\bm{n_2}, \bm{n_1}) \approx \frac{x\expf{x}}{\expf{x}-1}\,\beta\left[\cos\Theta_2-\cos\Theta_1\right],
\end{equation}
%------------------------
where $\Theta_2$ and $\Theta_1$ are the angles between the direction of motion and directions $\bm{n_2}$ and $\bm{n_1}$, respectively. To estimate the typical magnitude of the considered effect, we will use directions exactly along ($\Theta_2=0$) and opposite ($\Theta_1=\pi$) to the axis of the CMB dipole, which gives $\left[\cos\Theta_2-\cos\Theta_1\right] = 2$. 

%------------------------
\begin{figure}
	\begin{center}
		\includegraphics[width=\columnwidth]{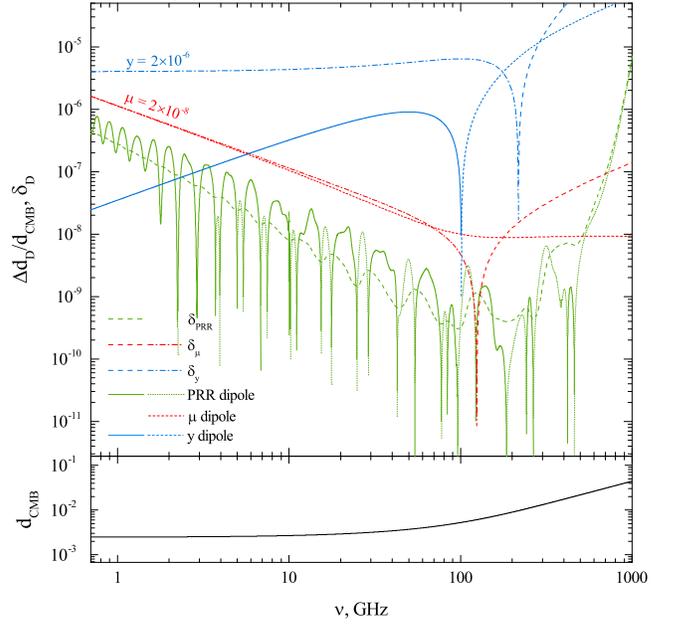}
		\caption{Top panel: The relative distortions of the CMB monopole, $\delta_{\rm D}$, and the relative distortions of the CMB dipole, ${\Delta d_{\rm D} / d_{\rm CMB}}$, as functions of frequency. Green, red, and blue lines correspond to the PRR-, $\mu$-, and $y$-distortions, respectively. For each type of distortion, dashed lines indicate the positive part of the monopole distortion, $\delta_{\rm D}$, and dashed-dotted the negative part. Solid lines indicate the positive part of the dipole distortion, ${\Delta d_{\rm D} / d_{\rm CMB}}$, and short dashed the negative. We adopted  $y=2\times10^{-6}$ and $\mu=2\times10^{-8}$ as the fiducial values for the $y$- and $\mu$-distortions, respectively. Bottom panel: the relative CMB dipole anisotropy, $d_{\rm CMB}$ (calculated using mutually opposite directions collinear to the CMB dipole axis), as a function of frequency.}
		\label{result}
	\end{center}
\end{figure}
%------------------------

To isolate the contributions arising only from the distortions of the CMB monopole, it is useful to consider
%------------------------
\begin{equation}
{\Delta d_{\rm D} \over d_{\rm CMB}}= {d_{{\rm CMB}+{\rm D}} \over  d_{\rm CMB}} -1,
\end{equation}
%------------------------
which defines the relative change of $d(\nu, \bm{n_2}, \bm{n_1})$ caused by the distortion of type "D", where "D" takes the values "PRR", "$y$", and `$\mu$'. 
It can be shown that the quantity $\Delta d_{\rm D}/d_{\rm CMB}$ is described by the following formula: 
%------------------------
\begin{equation}
{\Delta d_{\rm D} \over d_{\rm CMB}} \approx {\alpha_{\rm D} \over \gamma_{\rm CMB}}\delta_{\rm D}, 
\label{Dd_d_results}
\end{equation}
where 
\begin{equation}
\delta_{\rm D}(\nu)= {\eta_{\rm D, m}(\nu) \over \eta_{\rm CMB, m}(\nu)}
\end{equation} is the relative distortion of the CMB monopole spectrum with  
local spectral index $\alpha_{\rm D}=-\partial \ln \delta_{\rm D}/\partial\ln\nu$, and $\gamma_{\rm CMB}=-\partial \ln \eta_{\rm CMB, m}/\partial\ln\nu = x\expf{x}/(\expf{x}-1)$ 
is the local spectral index of the occupation number $\eta_{\rm CMB, m}$. It should be noted that $\alpha_{\rm D}=\left(\gamma_{\rm D}-\gamma_{\rm CMB}\right)$, where 
$\gamma_{\rm D}=-\partial \ln \eta_{\rm D, m}/\partial\ln\nu$ is the local spectral index of the occupation number $\eta_{\rm D, m}$. 
Note also that the spectral indices, $\gamma$, are defined corresponding to $\sim \nu^{-\gamma}$, such that positive values of $\gamma$ correspond to a decreasing spectrum.

Both $\Delta d_{\rm D} / d_{\rm CMB}$ and $d_{\rm CMB}$ are presented in Fig.~\ref{result}.  The dependence of $\Delta d_{\rm PRR}/d_{\rm CMB}$ on observation frequency shows quasi-oscillations originating from line features in the PRR spectrum. The amplitude of these quasi-oscillations ranges from $\pot{5.1}{-10}$ (at $\nu\simeq 228.5$~GHz) to $\pot{5.1}{-7}$ (at $\nu\simeq1.07$~GHz), i.e. 
$\Delta d_{\rm PRR}/d_{\rm CMB}$ increases at low frequencies. 
The value of $\Delta d_{y}/d_{\rm CMB}$ ranges from {$\pot{3.5}{-8}$} at 1 GHz to {$\pot{-4.2}{-5}$} at 700~GHz with a local maximum at the level of {$\pot{9.1}{-7}$} at 50.2~GHz, and null (change of sign) at 101.1 GHz. 
Thus, in contrast to $\Delta d_{\rm PRR}/d_{\rm CMB}$, the relative distortion anisotropy due to the average $y$-distortion, $\Delta d_{y}/d_{\rm CMB}$, tends to zero at lower frequencies. 
This implies that at $\nu\lesssim 6$ GHz, the signal due to the PRR becomes larger than the relative distortion of the CMB dipole caused by the $y$-distortion. The value of $|\Delta d_{\rm \mu}/d_{\rm CMB}|$ ranges from $\pot{9.3}{-9}$ at 700~GHz to $\pot{1.1}{-6}$ at 1~GHz showing a monotonic increase at decreasing frequency.

The following aspects should be emphasized.
\begin{itemize}
	\setlength\itemsep{-4pt}
	\item[a.] The relative distortion, $\Delta d_{\rm D} / d_{\rm CMB}$, is independent of the particular observation direction to leading order.
	\item[b.] A comparison of $\Delta d_{\rm PRR}/d_{\rm CMB}$ and $\delta_{\rm PRR}$ (the top panel of Fig.~\ref{result}) shows that the relative distortion of the CMB dipole can be slightly larger than the relative distortion of the CMB monopole due to the PRR. 
	\item[c.] At the same time, $\Delta d_{y}/d_{\rm CMB}$ is much smaller than $\delta_{y}$ at $\nu\lesssim 100$~GHz. Thus, the amplitude of $|\Delta d_{\rm PRR}/d_{\rm CMB}|$ is larger than 
	$|\Delta d_{y}/d_{\rm CMB}|$ at $\nu\lesssim 6$ GHz (as mentioned above) despite the fact that $|\delta_{\rm PRR}|\ll|\delta_{y}|$ in most part of {the} considered range 
	(aside from around 217 GHz -- the null-point of $\delta_{y}$). 
	\item[d.] At low frequencies, the relation $\Delta d_{\rm \mu}/d_{\rm CMB}\simeq \delta_{\rm \mu}$ is valid. 
\end{itemize}

The statement (a) directly follows from Eq.~\eqref{Dd_d_results}. Similarly, (b)$-$(d) can be understood by using the analytical approximation given above. 
To show this, let us introduce an {\it amplification} coefficient (in the same sense as in \citealt{Kholupenko2015}) 
\begin{equation}
C_{\rm D}={\Delta d_{\rm D} / d_{\rm CMB}\over \delta_{\rm D}\left(\nu\right)}\approx \frac{\alpha_{\rm D}}{\gamma_{\rm CMB}},
\label{ampl_coeff1}
\end{equation}
which characterizes how much the relative distortion of the dipole is larger than the corresponding relative contribution to the monopole. In the Rayleigh-Jeans part of the CMB, $\gamma_{\rm CMB}\simeq 1$ and hence 
$C_{\rm D}\simeq \alpha_{D}=\gamma_{\rm D}-1$. 

For the PRR, the amplification coefficient, $C_{\rm PRR}$, achieves values up to 6 (at the frequencies near 22, 26, and 107 GHz). For the $y$-distortion the amplification coefficient tends to zero ($C_{y}\rightarrow 0$) at low frequencies ($x\ll 1$) since the $y$-distortion of the CMB monopole has asymptotic $\delta_{y}\simeq -2y$ at $x \rightarrow 0$, and, correspondingly, $\alpha_{y}\rightarrow 0$. For the $\mu$-distortion, the amplification coefficient tends to unity ($C_{\rm \mu}\rightarrow 1$) at low frequencies ($\mu \ll x\ll 1$), since $\delta_{\rm \mu}\simeq -\mu/x$ at $x \rightarrow 0$ (but $x\gg \mu$) and, correspondingly, $\alpha_{\rm \mu}\rightarrow 1$. Note that the change of sign of $\Delta d_{\rm D} / d_{\rm CMB}$ (see Fig. \ref{result}) occurs at points of $\delta_{\rm D}$-extrema, where $\alpha_{\rm D}=0$. 

\begin{figure}
\begin{center}
\includegraphics[width=\columnwidth]{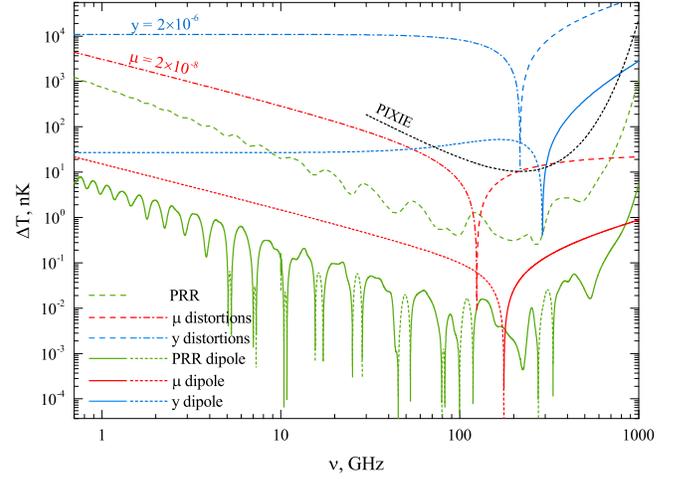}
\caption{Monopole temperature deviations, $\Delta T_{\rm D}$, and the dipole temperature differences, $\Delta T^{\rm d}_{\rm D}$ (in nK), arising from different types of relic CMB distortions, as functions of frequency.  
Green, red, and blue lines correspond to the PRR-, $\mu$-, and $y$- distortions, respectively. 
For each type of distortion, the dashed lines indicate the positive part of the monopole temperature deviations, $\Delta T_{\rm D}$, the dashed-dotted the negative. The solid lines indicate the positive part of the dipole temperature differences, $\Delta T^{\rm d}_{\rm D}$, and the short dashed the negative.  
The black dashed line correspond to the monopole sensitivity of PIXIE, $\Delta I_{\nu} = 5$ Jy\,sr$^{-1}$ (the expected limit for dipole is $\sqrt{2}$ lower).  All calculations of dipole temperature differences were performed for mutually opposite directions collinear to the CMB dipole axis. We adopted $y=2\times10^{-6}$ and $\mu=2\times10^{-8}$ as fiducial values for the $y$- and $\mu$-distortions, respectively.
}
\label{intensity}
\end{center}
\end{figure}

In Fig.~\ref{intensity}, we show the effective temperature differences from the individual distortion contributions {to} the dipole spectrum, $\Delta T^{\rm d}_{\rm D}\left(\nu, \Theta_1, \Theta_2\right)\approx T_0 \left[\eta_{\rm D} \left(\nu, \Theta_2\right) - \eta_{\rm D} \left(\nu, \Theta_1\right)\right]/G(x)$,
and compare it with the associated effective temperature difference of the monopole, 
$\Delta T_{\rm D}\left(\nu\right)\approx T_0 \,\eta_{\rm D, m} \left(\nu\right)/G(x)$.
We note that in contrast to $\Delta d_{\rm D} / d_{\rm CMB}$, the effective temperature dipole anisotropy, $\Delta T^{\rm d}_{\rm D}$, depends on direction, 
namely, it is directly proportional to $\left(\cos \Theta_2 - \cos \Theta_1\right)$. 

The PRR monopole temperature deviation shows an increase at low frequencies with marked quasi-oscillations originating from the wide overlapped lines in the PRR spectrum. This temperature deviation has a value in the range from $0.3$~nK at 270~GHz to $755$~nK at 1~GHz. The $y$-distortion monopole temperature deviation ranges from $-11$~$\mu$K at 1~GHz to $45$~$\mu$K at 700~GHz, a signal that will be easily detected using PIXIE. The $\mu$-distortion monopole temperature deviation ranges from $-3$~$\mu$K at 1~GHz to $20$~nK at 700~GHz.

\begin{table*}
	\begin{center}
		\begin{tabular}{cccccccc}
			\hline
			distortion & monopole/dipole & behavior & frequency of & min $\Delta T$, nK & frequency of & max $\Delta T$, nK \\
			& & & minimum, GHz & & maximum , GHz & \\
			\hline
			$y-$ & monopole & monotonic & 1 & -11$\times10^3$  & 700  & 45$\times10^3$ \\
			& dipole & monotonic & 1 & -27  & 700  & 1.1$\times10^3$ \\
			$\mu-$ & monopole & monotonic & 1 & -3$\times10^3$  & 700  & 20 \\
			& dipole & monotonic & 1 & -15  & 700  & 0.6 \\
			PRR & monopole & quasi-oscillating & 270 & 0.3  & 1  & 755 \\
			& dipole & quasi-oscillating & 89 & 7.3$\times10^{-3}$  & 1.07  & 5.1 \\
			\hline
		\end{tabular}
		\caption{The comparison of the monopole and dipole temperature differences for the different types of distortions in frequency range 1-700 GHz. Dipole temperature difference we calculated using mutually opposite direction collinear to CMB dipole axis. We adopted $y=2\times10^{-6}$ and $\mu=2\times10^{-8}$ as fiducial values for $y-$ and $\mu-$ distortions, respectively.}
		\label{comparison}
	\end{center}
\end{table*}

The PRR dipole anisotropy temperature difference also shows marked quasi-oscillations due to the lines in the PRR spectrum.
The amplitude of these quasi-oscillations has a value in the range from $\pot{7.3}{-3}$\,nK at $\nu\simeq 89$\,GHz 
to $5.1$\,nK at $\nu \simeq 1.07$\,GHz. 
The $y$-distortion dipole anisotropy temperature difference ranges from $-27$\,nK at 1\,GHz to $1.1\,\mu$K at 700\,GHz. 
The $\mu$-distortion dipole anisotropy temperature difference reaches $-15$\,nK at 1\,GHz to $0.6$\,nK at 700\,GHz.
Table~\ref{comparison} summarizes the comparison between the monopole and dipole signals for the various types of distortions.

From Fig.~\ref{intensity}, one can see that a detection of the distortion of the CMB (monopole) spectrum requires measurements of the signal in several frequency channels with the following values of absolute temperature sensitivity: 
(1) for PRR: $100$\,nK  at frequencies $\lesssim 4$ GHz, $10$\,nK at  $\lesssim 20$ GHz, $0.3$\, nK at $\lesssim 700$\,GHz (in whole considered range);
(2) for $y$-distortion: $4$\,$\mu$K  at frequencies $\lesssim 165$ GHz and $\gtrsim 265$\,GHz; 
(3) for $\mu$-distortion: $1$\,$\mu$K  at frequencies $\lesssim 3$\,GHz, $100$\, nK at  $\lesssim 25$\,GHz, $10$\,nK at $\lesssim 90$\,GHz and $\gtrsim 200$\,GHz.

On the other hand, from Fig.~\ref{intensity} one can also see that a detection of the dipole anisotropy due to the different distortions requires measurements of the signal in several frequency channels with the following values of absolute temperature sensitivity: 
(1) for PRR: $1$\,nK  at frequencies $\lesssim 4$\,GHz, $0.1$\,nK at  $\lesssim 20$\,GHz, and $10^{-3}$\,nK at $\lesssim 700$\,GHz; 
(2) for $y$-distortion: $4$\,nK  at frequencies $\lesssim 700$\,GHz ({excluding the} narrow band 285$-$293\,GHz around null of effect at 289.6 GHz); 
(3) for $\mu$-distortion: $1$\,nK  at frequencies $\lesssim 15$ GHz, $0.1$\,nK at  $\lesssim 100$\,GHz, and $\gtrsim 270$\,GHz.
Thus, observations of CMB dipole distortions require a sensitivity two to three orders of magnitude better than observations of corresponding distortions of CMB (monopole) spectrum in comparable ranges. The achievement of such an unprecedented level of sensitivity {is a challenging technological problem} for a future experiment observing the distortion of the CMB dipole. Also, the level of foreground contamination to the dipole spectrum with associated motion-induced effects has to be considered more carefully.

\section{Conclusion} 
The distortions of the dipole anisotropy due to several relic CMB distortions were studied. The temperature differences arising from the PRR-, $y$-, and $\mu$-dipoles depend on the frequency and observation directions. The relative change of the CMB dipole due to the PRR-, $y$-, and $\mu$-distortions is independent of the observation direction but depends on frequency. In the most promising range for observations of the PRR- and $\mu$-dipoles, $\nu\simeq1-20\,$GHz, the absolute temperature differences arising from these {signals} have values in the range $\simeq 0.1 -10\,$nK for observations in mutually opposite directions along the CMB dipole axis. The relative changes of CMB dipole due to the PRR and $\mu$-distortion in this range have values in the range $10^{-8}$ -- $10^{-6}$. In the most appropriate range for observations of the $y$-dipole, $\nu\simeq 20 - 700\,$GHz, the absolute value of the temperature difference ranges from 28\,nK to $1.1\,\mu$K, which could be detected at $\simeq 5\sigma$ with PIXIE. The relative change in the CMB dipole due to $y$-distortions in this range \textbf{is} $10^{-7}$ -- $10^{-5}$. 

The considered effect provides an additional method for detecting the relic CMB distortions. The main idea of the method is using an anisotropy technique that is already well developed for the measurements of the CMB anisotropy, with different systematics. One advantage of this method is that it does not require absolute, but only precise inter-channel calibration to observe spectral features, directly using the CMB sky as reference. This is especially important for the frequency range $\lesssim20$ GHz, where absolute calibration is difficult due to the size of cryogenic systems. The PRR dipole anisotropy has a unique spectral shape that is related to the initial PRR monopole spectrum. This can help experimentalists to identify reliably the PRR dipole anisotropy. The method requires absolute temperature sensitivity of at least 0.1$-$1\,nK. A sensitivity of 100\,nK has already been achieved using ground-based instruments (e.g., the Atacama Cosmology Telescope (ACTPol experiment; \citep{ACT2014}), the South Pole Telescope (SPT-POL experiment; \citep{SPTPOL2012}) and BICEP2 \citep{Bicep2014}, which has a sensitivity of 85\,nK). However, these are not well suited for measurements of the dipole spectrum, since it is difficult to access large scales because of atmospheric fluctuations. The general consensus is that these kind of distortion measurements will need to be performed from space or possibly with the advent of the next generation of the bolometer telescopes (e.g. PIXIE, \citealt{PIXIE}; PRISM, \citealt{PRISM}; and CMB Stage-IV, \citealt{Chang2013}). At low frequencies ($\lesssim20$\,GHz), the receivers become larger with increasing wavelength, so it is difficult to observe with sensitive spectrometers in space. One possible resolution is the measurements using balloon-borne experiments, like ARCADE 2 \citep{ARCADE2}, or finding a new path to make observations from the ground.}

As mentioned above, the motion-induced dipole signal related to the average $y$-distortion may be detected with PIXIE. Since the signal has a spectrum that differs from the usual $y$-distortion, it cannot be mimicked by a dipolar modulation of the number of clusters (or generally scatters) across the sky. Thus, the distortion of the dipole can in principle be used to independently confirm that the CMB temperature dipole is caused primarily by our motion, thereby placing limits on a primordial temperature dipole and large-scale perturbations.

\vspace{2mm}{\footnotesize {\rm Acknowledgments}
J.C. thanks Yacine Ali-Ha{\"i}moud for useful discussions. S.A.B., A.V.I. and D.A.V. are grateful to Russian Science Foundation grant 14-12-00955. E.E.K. thanks RFBR grant 13-02-12017-ofi-m. J.C. is supported by the Royal Society as a Royal Society University Research Fellow at the University of Cambridge, UK.

\bibliography{Relic_bg_Dipole}

\end{document}